\documentclass[conference]{IEEEtran}
\usepackage{color,graphicx}
\usepackage{url}
\usepackage{amsmath}
\usepackage{algorithmic}
\usepackage{algorithm}
\usepackage{epstopdf}
\usepackage{epsfig}
\usepackage{cite}
\usepackage{subcaption}

\ifCLASSINFOpdf
\else
\fi

\makeatletter

\newcommand{\Rmnum}[1]{\expandafter\@slowromancap\romannumeral #1@}
\makeatother
\begin{document}
\title{“Association fairness” in Wi-Fi and LTE-U coexistence}
\vspace{-0.5cm}
\author{Vanlin Sathya$^\dag$, Morteza Mehrnoush$^*$, Monisha Ghosh$^\dag$, and Sumit Roy$^*$
\IEEEauthorblockN{}
\IEEEauthorblockA{$^\dag$University of Chicago, Illinois, USA. $^*$University of Washington, Seattle, USA. \\
{Email: vanlin@uchicago.edu, mortezam@uw.edu, monisha@uchicago.edu, sroy@u.washington.edu.}}
}
\maketitle

\begin{abstract}
In this paper we address the issue of \textit{association fairness} when Wi-Fi and LTE unlicensed (LTE-U) coexist on the same channel in the unlicensed 5 GHz band. Since beacon transmission is the first step in starting the association process in Wi-Fi, we define \textit{association fairness} as how fair LTE-U is in allowing Wi-Fi to start transmitting beacons on a channel that it occupies with a very large duty cycle. According to the LTE-U specification, if a LTE-U base station determines that a channel is vacant, it can transmit for up to 20 ms and turn OFF for only 1 ms, resulting in a duty cycle of 95\%. In an area with heavy spectrum usage, there will be cases when a Wi-Fi access point wishes to share the same channel, as it does today with Wi-Fi. We study, both theoretically and experimentally, the effect that such a large LTE-U duty cycle can have on the association process, specifically Wi-Fi beacon transmission and reception. We demonstrate via an experimental set-up using National Instrument (NI) USRPs that a significant percentage of Wi-Fi beacons will either not be transmitted in a timely fashion or will not be received at the LTE-U BS thus making it difficult for the LTE-U BS to adapt its duty cycle in response to the Wi-Fi usage. Our experimental results corroborate our theoretical analysis. We compare the results with Wi-Fi/Wi-Fi coexistence and demonstrate that LTE-U/Wi-Fi coexistence is not “fair” when it comes to initial association since there is a much larger percentage of beacon errors in the latter case. Hence, the results in the paper indicate that in order to maintain association fairness, a LTE-U BS should not transmit at such high duty cycles, even if it deems the channel to be vacant.\\

\end{abstract}

%

\vspace{-0.3cm}
\section{Introduction}
Driven by increasing user demands for greater bandwidth
and limited licensed spectrum, cellular systems will soon be
deployed in the 5 GHz unlicensed bands, which are primarily
used by Wi-Fi today. Two systems are under consideration for this deployment: (i) LTE-LAA~\cite{3GPP, kini2016wi}, which is being developed by the 3GPP standardization body and uses listen-before-talk (LBT), which is similar to carrier sense multiple access with collision avoidance (CSMA/CA) used by Wi-Fi and (ii) LTE unlicensed (LTE-U~\cite{LTEU, pang2016wi}), which is being developed by an industry consortium (LTE-U Forum) and employs a much simpler, but potentially more harmful to Wi-Fi, coexistence mechanism which depends on duty-cycling along with a ``light'' sensing technique called Carrier Sense Adaptive Transmission (CSAT) that adapts the duty cycle depending on the perceived Wi-Fi usage at the LTE-U base station (BS). In this paper we focus on LTE-U in the following specific scenario: according to the LTE-U specification, if a channel is ``vacant'', i.e. no Wi-Fi is detected, a LTE-U BS can transmit for a maximum 20 ms ON time and a minimum 1 ms OFF time thus leading to a 95\% duty cycle~\cite{forum}. If a Wi-Fi access point (AP) now wishes to also share this channel with LTE-U, it has to begin by transmitting beacons, which are also subject to CSMA/CA. However with such a large duty cycle, and only 1 ms of OFF time, it is unclear as to how successful it will be in setting up its beacon transmissions, which is a necessary prerequisite before association with other Wi-Fi devices can take place. We call this criterion \textit{association fairness}, to distinguish it from \textit{throughput fairness} in which the LTE-U and Wi-Fi share the channel fairly to achieve the same system throughput. While there have been a number of papers investigating \textit{throughput fairness} as a function of detection threshold and duty-cycle when Wi-Fi and LTE-U coexist~\cite{cano2015unlicensed,chai2016lte,almeida2013enabling,chen2016optimizing}, this particular issue of \textit{association fairness} has not received much attention.

\par In \cite{G_whitepaper}, an initial investigation by Google into the issue of coexistence between LTE-U and Wi-Fi is summarized. LTE-U can lead to long consecutive strings of missed beacons for Wi-Fi clients that give up on beacon reception after some pre-defined time, thus making dis-association even more likely than the average  beacon delay time indicates. The association fairness problem is briefly described in this paper, but a comprehensive data analysis is missing. 



In this paper, we study specifically the association fairness issue theoretically as well as with an experimental test-bed using National Instrument (NI) USRPs that transmit a LTE-U signal, real Wi-Fi APs and laptops with Wireshark~\cite{wiki:xxx} installed to capture beacon transmissions and receptions, with the goal of quantifying the effect of LTE-U duty cycle and OFF time on Wi-Fi beacon transmission and reception. The beacon drop probability and expected delivery time is modeled analytically and corroborated via NI experimental testbed for validating the results. Our results indicate that even when the channel is not occupied by a Wi-Fi AP, LTE-U should not use its maximum allowed duty cycle of 95\% in order to allow Wi-Fi APs to begin the association for sharing the channel.

 
 \vspace{-0.2cm}
\section{Background}
In this section, we present the system model that we consider in this paper, a brief background on Wi-Fi beacon transmissions and the LTE-U signaling scheme.
 \vspace{-0.2cm}
 \begin{figure}[htb!]
\begin{center}
\includegraphics[totalheight=2.7cm,width=6.8cm]{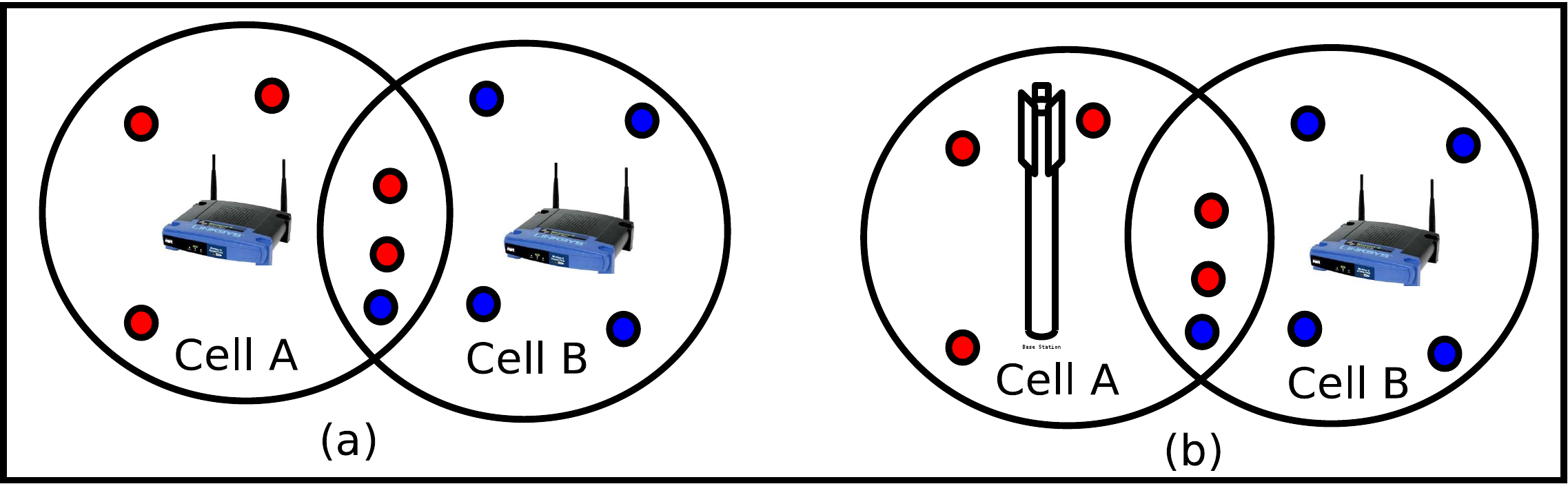}
\end{center}
\vspace{-0.2cm}
\caption{(a) Cell A and Cell B use Wi-Fi, and (b) Cell A switches to LTE-U}
\vspace{-0.5cm}
\label{wifi}
\end{figure}

\subsection{Coexistence System Model}

We consider LTE-U/Wi-Fi coexistence in the unlicensed 5 GHz band. The LTE-U transmissions in the unlicensed band are only on the downlink, with all uplink traffic (acknowledgements and control) being transmitted on a licensed channel. We will compare a LTE-U/Wi-Fi coexistence scenario with a Wi-Fi/Wi-Fi coexistence scenario to evaluate the association fairness in the two cases. Figure~\ref{wifi} (a) shows the deployment configuration being considered with both Cell A and Cell B using Wi-Fi, where the users associated with Cell A and Cell B are denoted by red and blue respectively. Figure~\ref{wifi} (b) shows Cell A switching to LTE-U. We will evaluate the association issue by varying the duty cycle for the LTE-U transmission and examining the effect on Wi-Fi beacon transmission and reception.
\vspace{-0.1cm}
\subsection{Wi-Fi Association process}
\label{subsec: Assoc}

Wi-Fi allows two kinds of beacon scanning: passive and
active. In passive scanning, the Wi-Fi client passively listens to a beacon transmission from the AP. When it is successfully received, the client gathers the BSS information and initiates the association process with the AP. In active scanning, the client actively requests a beacon by broadcasting a probe request packet. The AP then replies to the client by sending a probe response packet (\emph{i.e.,}unicast packet) which is similar
to a beacon packet. Most Wi-Fi APs support both active and
passive scanning modes, i.e. both beacon and probe response
packets are transmitted by the AP during the association
process. In this paper we analyze passive scanning only, which depends on reliable transmission and reception of beacon frames.

\vspace{-0.1cm}
\subsection{Wi-Fi Beacon transmission}\label{s1}

To announce its presence a Wi-Fi AP periodically transmits beacon frames. The beacon frame consists of the service set identifier (SSID) which is used to identify its infrastructure basic service set (BSS), and its network capability information. Other information sent in a beacon frame includes time stamp, beacon interval, supported data rates, traffic indication map (TIM), BSS load, QoS capability, etc. Beacon frames are sent periodically at an interval defined in Time Units (TU), which is typically configured in the AP as 100 TU (102.4 ms). The size of a beacon frame varies from 60 to 450 bytes, depending on the information being transmitted. Beacon frames are also transmitted with CSMA/CA~\cite{std80211}, i.e. the AP needs to check for the availability of the channel before sending beacon packets. If the channel is sensed to be idle, the AP will perform a random back-off with the minimum contention window (CWmin). During back-off, if the channel is sensed to be busy, the AP defers until the channel is again sensed to be idle. Since there is no acknowledgment (ACK) in beacon transmissions, the AP just selects the random back-off based on CWmin. If the beacon packet is transmitted, but not received at a particular location, it will not be re-transmitted because it is a broadcast packet and there is no ACK.



\vspace{-0.1cm}
\subsection{LTE-U}
LTE-U uses the basic LTE frame structure, i.e. a frame length of 10 ms. There is no LBT or CSMA/CA employed before every data transmission. Instead, a duty-cycling approach is used where the duty cycle is determined by perceived Wi-Fi usage at the LTE-U BS, using carrier sensing. If a single Wi-Fi AP is detected, the duty cycle is set at 50\%. Once the duty cycle is set, the LTE-U BS is allowed to transmit data over the unlicensed channel during the ON state without LBT. Conversely, the LTE-U BS is expected to not operate on the unlicensed channel during the OFF state. According to the LTE-U Forum specification, the duration of ON state should be more than 4 ms and less than 20 ms. Also, the duration of the OFF state should be at least 1 ms. Hence, if no active Wi-Fi AP is detected, a LTE-U BS can start transmissions with a ON time of 20 ms and a OFF time of 1 ms, resulting in a duty cycle of 95\%. During the ON period, downlink transmissions to UEs are scheduled by the LTE-U BS, unlike Wi-Fi where each transmission has to be preceded by a CSMA/CA process. 



\vspace{-0.1cm}
\subsection{Association Process in LTE-U Wi-Fi Coexistence}

\begin{figure}[!htb]
  \includegraphics[height=4cm,width=9cm]{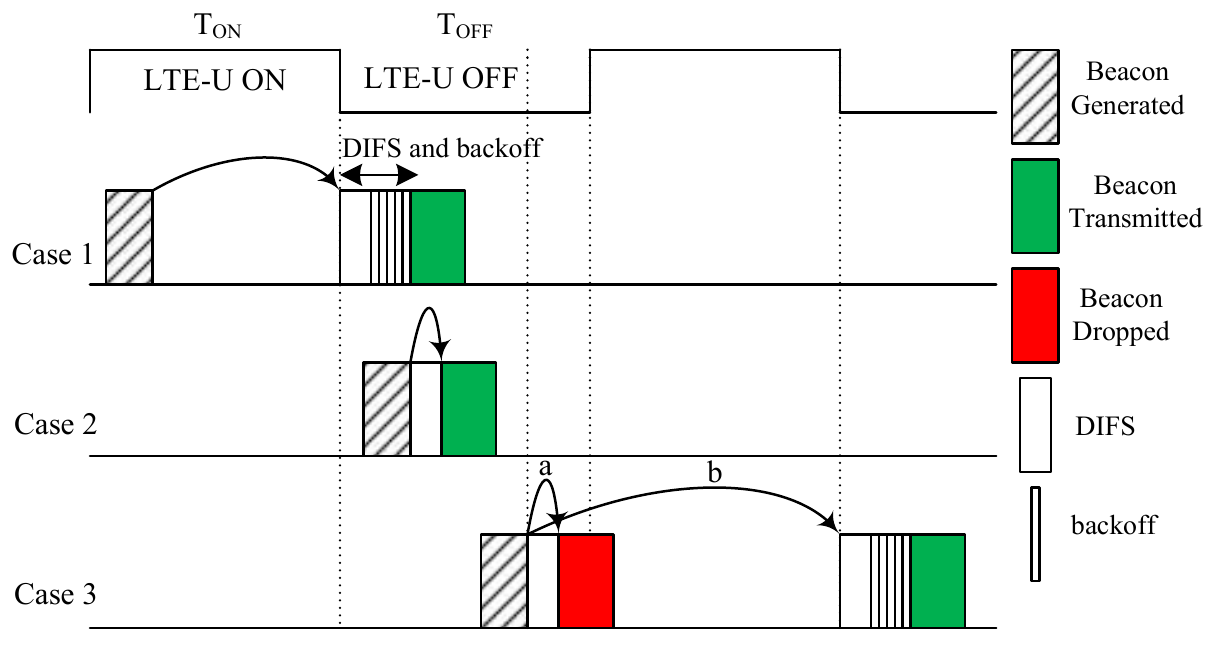}
  \vspace{-0.5cm}
  \caption{Possible cases for beacon transmission and reception during one ON/OFF LTE-U cycle}
   \vspace{-0.3cm}
  \label{bdrop}
\end{figure}

The beacon transmission and reception process in LTE-U/Wi-Fi coexistence is illustrated in Figure~\ref{bdrop} with an example LTE-U duty cycle of 50\%.

In Case 1, the beacon is generated during the LTE-U ON period, with a periodicity determined by the AP with a nominal value of 102.4 ms. Since the channel is busy due to the LTE-U ON time, the Wi-Fi AP waits until the end of the ON period, senses the channel for a time equal to DCF interframe space (DIFS), selects a random back-off and transmits the beacon if the channel is idle. Since the minimum length of the OFF period is 1 ms which is larger than the beacon transmission time of 427$\mu s$ as given in Table~\ref{tab: beaconpara} plus the DIFS and back-off time, overlap with the second ON period does not happen and the beacon is transmitted and received successfully. 

In Case 2, the beacon is generated during the LTE-U OFF period, the channel is idle, the AP performs DIFS, then random back-off and transmits. In this case also the beacon transmission time plus DIFS, and back-off time is such that the beacon does not overlap with the second ON period and the beacon is transmitted and received successfully. 
 \vspace{-0.4cm}
In Case 3, the beacon is generated in the LTE-U OFF period and the DIFS sensing shows that the channel is idle. Now, two situations are possible: (a) the random back-off reaches zero and the AP transmits the beacon but partially or completely overlaps with the subsequent LTE-U ON period, or (b) before finishing the random back-off the subsequent LTE-U ON period starts which causes the AP to wait till the end of the LTE-U ON period and continue the back-off afterwards. The length of the beacon transmission time is smaller than the minimum OFF period (1 ms), so the generated beacon generally is either transmitted in the current period or differed to the start of the next OFF period.

In Case 1 and Case 3b above, the nominal inter-beacon transmission interval of 102.4 ms will increase and in Case 3a, the beacon will be transmitted but may not be received successfully since while some errors in the beacon frame can be corrected by the error correction coding, if a sufficiently large portion of the beacon frame is interfered with by the LTE-U transmission, it will not be received correctly. Our theoretical analysis and experimental study will quantify both these types of events as a function of duty cycle and period.

The above cases assumed that during the association process only beacons were being transmitted (no data or other association frames), but as explained in Section \ref{subsec: Assoc} most Wi-Fi APs will respond to probe requests from clients with a probe response, which is an unicast packet with a corresponding ACK. If the probe response transmission is unsuccessful, the AP will double its contention window on the next transmission. These packets will also be transmitted during the LTE-U OFF period and hence will have an impact on beacon frame transmission and reception as well.

 \vspace{-0.3cm}
\begin{table}[htb!]
\caption{Beacons Transmission Parameters}
\vspace{-0.2cm}
\centering
\begin{tabular}{|p{5cm}| p{2cm}|}
\hline
\bfseries Parameter &\bfseries Value \\ [0.4ex]
\hline\hline
Wi-Fi mode & IEEE 802.11 ac\\
\hline
DIFS & 34 $\mu s$\\ 
\hline
CWmin (W) & 16 \\
\hline
Beacon Frame Length & 305 bytes \\
\hline
Beacon Transmission Data Rate & 6 Mbps \\
\hline
Beacon transmission time ($T_b$) & 427 $\mu s$ \\
\hline
PHY preamble & 20 $\mu s$ \\
\hline
Time slot ($t_s$) & 9 $\mu s$ \\
\hline
\end{tabular}
\label{tab: beaconpara}
\end{table} 

For comparison with the above, let us consider the case of Wi-Fi/Wi-Fi coexistence. Assume that there is a single Wi-Fi AP on a channel with fully loaded traffic: this is analogous to the case we will consider later with LTE-U using a 95\% duty cycle. If a second Wi-Fi AP wishes to use this channel, perhaps because no other vacant channel is available for its use, it will commence by transmitting beacon frames advertising its presence. Since both Wi-Fi APs use CSMA/CA on beacon and data frames, the new AP will be able to access the channel “fairly” and begin transmitting beacons successfully. This is in contrast to the situation described above with LTE-U/Wi-Fi coexistence where beacons can get delayed or not received correctly, especially with a duty cycle of 95\% where the second Wi-Fi AP will initially have extremely limited access to start transmitting beacons since LTE-U does not back-off. There is a chicken and egg problem here in that LTE-U is supposed to reduce its duty-cycle when it detects Wi-Fi, but if Wi-Fi is unable to even start transmitting beacons, how will LTE-U detect Wi-Fi and scale back its duty cycle appropriately?
\vspace{-0.3cm}
\section{Beacon Drop Probability}\label{p32}

In this section, we calculate the beacon drop probability of Wi-Fi when it coexists with LTE-U to investigate how LTE-U affects the beacon frame transmission. We assume that the Wi-Fi and LTE-U are co-channel and that all  Wi-Fi stations (AP and clients) can perfectly detect the LTE-U in the ON period. As mentioned previously, beacon transmission follows the CSMA/CA protocol, however since it is a broadcast packet, there is no ACK and the random back-off is selected based on the minimum contention window (CWmin). The main reason that a Wi-Fi beacon once transmitted is not received is that the beacon frame overlaps with the LTE-U ON period after transmission.

To calculate the drop probability we consider Case 3a in Figure~\ref{bdrop}. We assume that if more than $P_o$\% of the beacon frame overlaps the subsequent LTE-U ON period, i.e. $\lceil P_o T_b/t_s\rceil$ time slots of the beacon frame overlap with second ON period where $t_s$ is slot time and $T_b$ is the beacon transmission time, the beacon would be dropped. This means that less than $P_o$\% overlap could be potentially corrected by the error correcting code. The probability that a beacon is generated at a time slot in the OFF period is:
\begin{equation}
\small
\begin{split}
P_s &= P(OFF)P(t_s|OFF)=\frac{T_{OFF}}{T_{ON}+T_{OFF}}\times \frac{t_{s}}{T_{OFF}}\\
&=\frac{t_{s}}{T_{ON}+T_{OFF}},
\label{ed: Ps}
\end{split}
\end{equation}
where $T_{ON}$ is the LTE-U ON, and $T_{OFF}$ is LTE-U OFF period. So, the beacon drop probability given the beacon transmission time and minimum overlap which causes the beacon to drop is calculated as:
\begin{equation}
\small
\begin{split}
P_{dt} &=P_s(\lceil (1-P_o)T_b/t_s\rceil).
\end{split}
\label{ed: Pt}
\end{equation}

From eq (\ref{ed: Pt}), we see that the drop probability is independent of the LTE-U duty cycle and depends only on the total length of one ON/OFF period i.e. the number of ON/OFF cycles which happen in one 102.4 ms beacon period or equivalently the number of ON edges in a 102.4 ms period. 
\vspace{-0.1cm}
\section{Additional Expected Delivery Time Calculation}
In this section, the beacon delivery time of Wi-Fi in coexistence with LTE-U is analyzed for both theoretical analysis and experimental calculation to investigate how the LTE-U duty cycle affects the Wi-Fi beacon transmission.
\vspace{-0.1cm}
\subsection{Theoretical Analysis}
To calculate the additional expected beacon delivery time of the beacon (this is the additional expected time from the beacon generation until the successful delivery time when coexisting with LTE-U), we have to capture the expected delay of Case 1, Case 2 and Case 3b in Fig.~\ref{bdrop}. The expected delay of beacon delivery in Case 1 is:
\begin{equation}
\small
E[T_1] = T_{ON}/2+\text{DIFS}+\frac{W-1}{2}t_s+T_b,
\label{ed: ET1}
\end{equation}
where the term $T_{ON}/2$ is the expected delay assuming the beacon generation in the LTE-U ON period follows the uniform distribution. The expected delay of beacon delivery in Case 2 is:
\begin{equation}
\small
E[T_2] = \text{DIFS}+T_b.
\label{ed: ET2}
\end{equation}

Similarly the expected delay of beacon delivery in Case 3b is:
\begin{equation}
\small
E[T_3] = \text{DIFS}/2+T_{ON}+\text{DIFS}+\frac{W-1}{2}t_s+T_b,
\label{ed: ET3}
\end{equation}
where the term $\text{DIFS}/2$ is the expected delay because the LTE-U ON starts in the DIFS sensing period. The total expected delivery time (in case the beacon is not dropped) is:
\begin{equation}
\small
\begin{split}
&E[T_{bt}] = P_bE[T_1]+(1-P_b) \\
& \times \left [\frac{T_{OFF}-(T_b+\text{DIFS})}{T_{OFF}}E[T_2]+\frac{\text{DIFS}}{T_{OFF}}E[T_3] \right],
\label{ed: ETt}
\end{split}
\end{equation}
where the $P_b=\frac{T_{ON}}{T_{ON}+T_{OFF}}$. In eq (\ref{ed: ETt}), the additional expected beacon delivery time directly depends on the LTE-U duty cycle ($P_b$) and the length of the ON period in each cycle ($T_{ON}$) unlike the drop probability  in eq. (2) which depends  only on $T_{ON}$.
\vspace{-0.1cm}
\subsection{Experimental Calculation} 
Let the Wi-Fi beacon generation times be: $b_1$, $b_2$, $b_3$, $\dots$, $b_N$. Nominally these are 102.4 ms apart, \emph{i.e.,} $b_{i+1}-b_{i}$ = 102.4. We cannot measure, experimentally, the actual beacon generation times, $b_i$. However, we can measure, using Wireshark, the beacon reception times: $t_1$, $t_2$, $t_3$, $\dots$, $t_N$. Let us assume there are no dropped beacons.
\par The theoretical analysis gives us an expression for E[$t_i$-$b_i$] in eq. (6), which experimentally can be calculated as follows.
\begin{equation}\label{d1}
\small
\begin{split}
\frac{1}{N} \sum_{i=1}^{N} (t_i-b_i) = & \frac{1}{N} [\sum_{i=1}^{N} t_i] - \frac{1}{N}
[ \sum_{i=1}^{N} (b_1 + 102.4 (i-1))]\\
= & \frac{1}{N} [\sum_{i=1}^{N} t_i] - b_1 - 102.4 \frac{N-1}{2}
\end{split}
\end{equation}
Hence, in order to match the experimental results with theory, we need to estimate the initial condition, which is the beacon generation
time of the first beacon, $b_1$. We will describe in the next section how we can do so experimentally.
\section{Experimental Results and Analysis} 
In this section, we first describe the experimental setup, followed by validation of the theoretical results from the previous section with careful experiments. The objective here is to accurately determine the number of successful Wi-Fi beacon transmissions and receptions for different duty cycles and OFF periods of the LTE-U so that we can characterize the association scenarios described in Section II.E. above. 
\vspace{-0.3cm}
\begin{figure}[htb!]
\begin{center}
\includegraphics[totalheight=2cm,width=7.8cm]{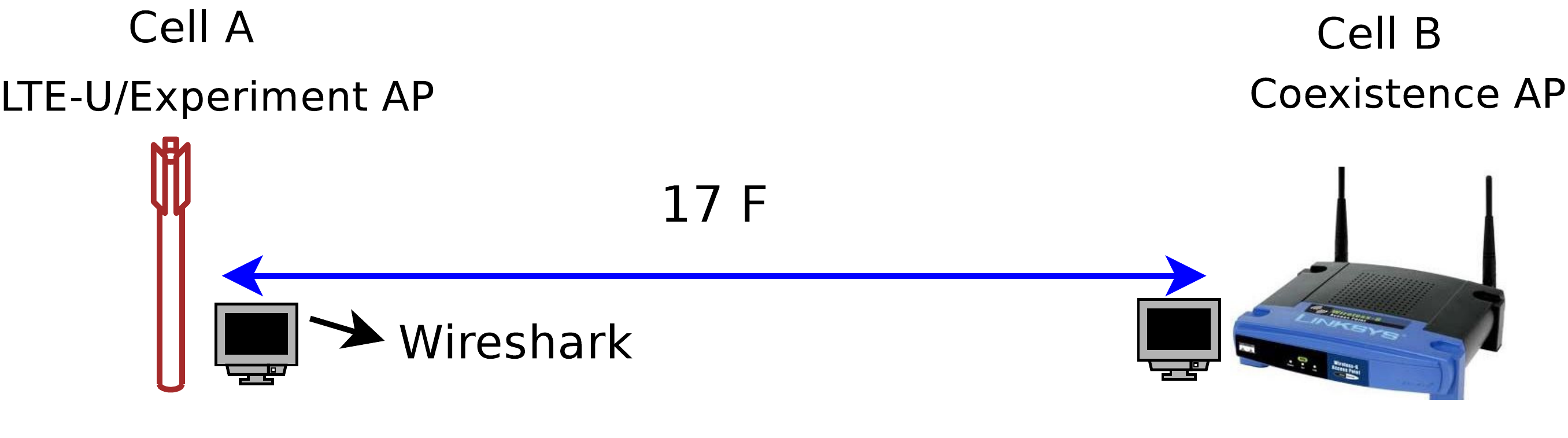}
\vspace{-0.2cm}
\caption{LTE Wi-Fi Co-existence Experimental Setup.}
 \vspace{-0.5cm}
\label{tbed}
\end{center}
\end{figure}

\begin{table}[htb!]
\label{table2}
\caption{Simulation parameters}
\vspace{-0.2cm}
\centering
\begin{tabular}{|p{4cm}| p{4cm}|}
\hline\bfseries
Parameter&\bfseries Value \\ [0.4ex]
\hline
Transmission Scheme & OFDM\\ 
\hline
Bandwidth & 20 MHz  \\
\hline
Operating frequency & 5.805 GHz   \\
\hline
Operating channel & 161\\
\hline
Transmission power for both LTE and Wi-Fi & 23 dBm \\ 
\hline
Traffic & Full Buffer (Saturation Case) \\
\hline
Transmission time interval & 1 ms \\
\hline
Time slot duration ($t_s$) & 9 $\mu s$ \\
\hline
Wi-Fi Energy Threshold & -82 dBm \\
\hline
Beacon transmission interval & 102.4 ms \\
\hline
Wi-Fi sensing protocol & CSMA/CA \\
\hline
Wi-Fi Antenna Type & MIMO \\
\hline
LTE-U Antenna Type & SISO \\
\hline
LTE-U ON/OFF period& 20 ms ON/OFF, 5 ms ON/OFF, 20 ms ON \& 5 ms OFF and 20 ms ON \& 1 ms OFF \\
\hline
LTE-U data and control channel & PDCCH and PDSCH \\
\hline
\end{tabular}
\label{table:Building1}
 \vspace{-0.5cm}
\end{table} 


 \vspace{-0.5cm}

\subsection{Experimental Setup}
The coexistence system model described in Section II is tested using a set-up based on a NI USRP and off-the-shelf Wi-Fi APs and client devices. The experiment was set up in an open lab environment where there are other Wi-Fi clients in the area that may be transmitting probe requests to the APs under test. We use the LTE coexistence framework~\cite{NI_whitepaper} and configure the NI USRP 2953R SDR to transmit a LTE-U signal and vary the ON and OFF times to obtain different duty cycles. Two Netgear APs are used as the Wi-Fi APs. Figure ~\ref{tbed} shows the experimental test-bed setup where Cell A is either the LTE-U BS or a Wi-Fi AP labeled as ``Experiment AP'' and Cell B is always Wi-Fi and is labeled ``Coexistence AP''. Cell B does not transmit any data, only transmits beacon frames (and probe responses, if clients in the vicinity transmit probe requests). The Wi-Fi APs and the NI LTE-U BS are provisioned to operate on the same unlicensed channel (Channel 161), and it was ensured that there are no other Wi-Fi APs on this channel. Since the NI LTE-U BS implementation does not implement CSAT, we measure how many Wi-Fi beacons are received at the LTE-U BS by using a laptop in monitor mode with Wireshark installed on it and placing it very close to the LTE-U BS. We will call the beacons received by this laptop as ``Wi-Fi beacons received at Cell A''. We also use Wireshark on another laptop in monitor mode at the same time near the transmitting Wi-Fi AP (Cell B), and call the beacons received by this laptop as ``Wi-Fi beacons transmitted by Cell B''. By comparing the sequence IDs of each beacon that is reported by the Wireshark at each laptop, we can compare how many beacons were successfully transmitted (from the laptop close to the Wi-Fi AP at Cell B) and received (from the laptop close to the Cell A). The LTE-U and Wi-Fi transmission characteristics and other parameters under study are summarized in Table~\ref{table:Building1}. We study Wi-Fi/Wi-Fi coexistence and LTE-U/Wi-Fi coexistence under different scenarios described below.
\vspace{-0.1cm}
\subsection{Theoretical Results and Comparison for Beacon Reception}\label{p32}

In this section we present experimental validation of the theoretical analysis of the beacon drop probability developed in Section III. In real deployments,  most Wi-Fi APs support both active and passive scanning modes, i.e., both beacon and probe response packets are transmitted by the AP during the association process. However, the generation of probe request packets (randomly broadcast by the Wi-Fi clients) is difficult to model theoretically. Hence, in order to verify the theoretical analysis, we disable\footnote{In order to turn off the probe request packets, we carried out the experiment late at night on the University of Chicago campus and verified from Wireshark that no probe request packets were generated.} the probe request packet for this particular experiment. The experimental result is calculated based on the number of received beacon packets out of 3000 transmitted packets, where the distance between the Wi-Fi AP and LTE-U is 17 feet and the Wi-Fi client is close to the AP.

Using the beacon parameters in Table~\ref{tab: beaconpara} and $P_o=0\%$ overlap, i.e. any overlap with a LTE-U ON period results in a beacon drop, the successful beacon reception probability (which is $1-P_{dt}$) is shown  in the theoretical column of Table~\ref{table:BeaconRe}. There is good agreement between the theory and the experiment, both of which demonstrate that the beacon drop probability is a function of the total period of one ON/OFF cycle of the LTE-U transmission. 
\vspace{-0.2cm}
\begin{table}[htb!]
\label{table2}
\caption{Theoretical and experimental Beacon Reception}
\vspace{-0.1cm}
\centering
\begin{tabular}{|c|c|c|c|}
\hline
Setup & Theoretical & Experimental \\ 
\hline\hline
$T_{ON}=T_{OFF}=$ 5ms & 0.9559 & 0.9626 \\ 
\hline
$T_{ON}=$ 20ms, $T_{OFF}=$ 1ms & 0.9794 & 0.9656 \\ 
\hline
$T_{ON}=T_{OFF}=$ 20ms & 0.9892 & 0.9700 \\ 
\hline
\end{tabular}
\label{table:BeaconRe}
\vspace{-0.3cm}
\end{table}

\subsection{Theoretical Results and Comparison for Beacon Delivery Time}

In this section we present experimental and theoretical expected delivery times as calculated in Section IV. Table~\ref{table:Deltime} shows the theoretical expected delivery time as calculated from eq. (\ref{ed: ETt}) and the experimental expected time of the beacon using eq. (\ref{d1}). In order to estimate the generation time of the first beacon $b_1$, we set up an experiment where the Wi-Fi AP begins transmitting beacons on a clean channel \emph{i.e.,} with no LTE-U or other interference, thus ensuring that the measured received time $t_i$ is approximately equal to the beacon transmitted time $b_i$. This gives a reference for $b_1$. We then turn on the LTE-U, and start measuring $t_i$ again. We use these measurements in eq. (\ref{d1}) to calculate the experimental values in Table~\ref{table:Deltime}. We observe that LTE-U with 20 ms ON/1 ms OFF causes the highest expected delivery time; while 5 ms ON/5ms OFF causes the smallest delay and that the overall trends remains the same in both theory and experiments.



\vspace{-0.2cm}
\begin{table}[!htbp]
\caption{Theoretical and experimental expected delivery time of the beacon.}
\vspace{-0.3cm}
\label{table:Deltime}
\begin{center}
\begin{tabular}{|c|c|c|}
\hline
Setup & Theoretical & Experimental \\ 
\hline\hline
$T_{ON}=T_{OFF}=$ 5ms & 1.82 ms& 1.01 ms \\ 
\hline
$T_{ON}=$ 20ms, $T_{OFF}=$ 1ms & 10.15 ms & 7.42 ms \\ 
\hline
$T_{ON}=T_{OFF}=$ 20ms & 5.59 ms & 4.24 ms \\ 
\hline
\end{tabular}
\end{center}
\vspace{-0.4cm}
\end{table}

\subsection{Performance comparison with LTE-U duty cycle of 95\%}\label{b1}
In this section we study the coexistence performance with a baseline experiment with a LTE-U duty cycle of 95\%, i.e. 20 ms ON/1 ms OFF as follows.
\begin{itemize}
\item \textbf{Step 1:} Only one Wi-Fi AP is deployed, Cell B in Figure ~\ref{tbed}.
 \item \textbf{Step 2:} Wi-Fi/Wi-Fi coexistence: both Cell A and Cell B use Wi-Fi.
 \item \textbf{Step 3:} LTE-U/Wi-Fi coexistence: Cell A switches to LTE-U and Cell B continues using Wi-Fi.
\end{itemize}

\vspace{-0.1cm}
\vspace{-0.4cm}
 \begin{figure}[!htb]
 \centering
  \includegraphics[height=5cm,width=9cm]{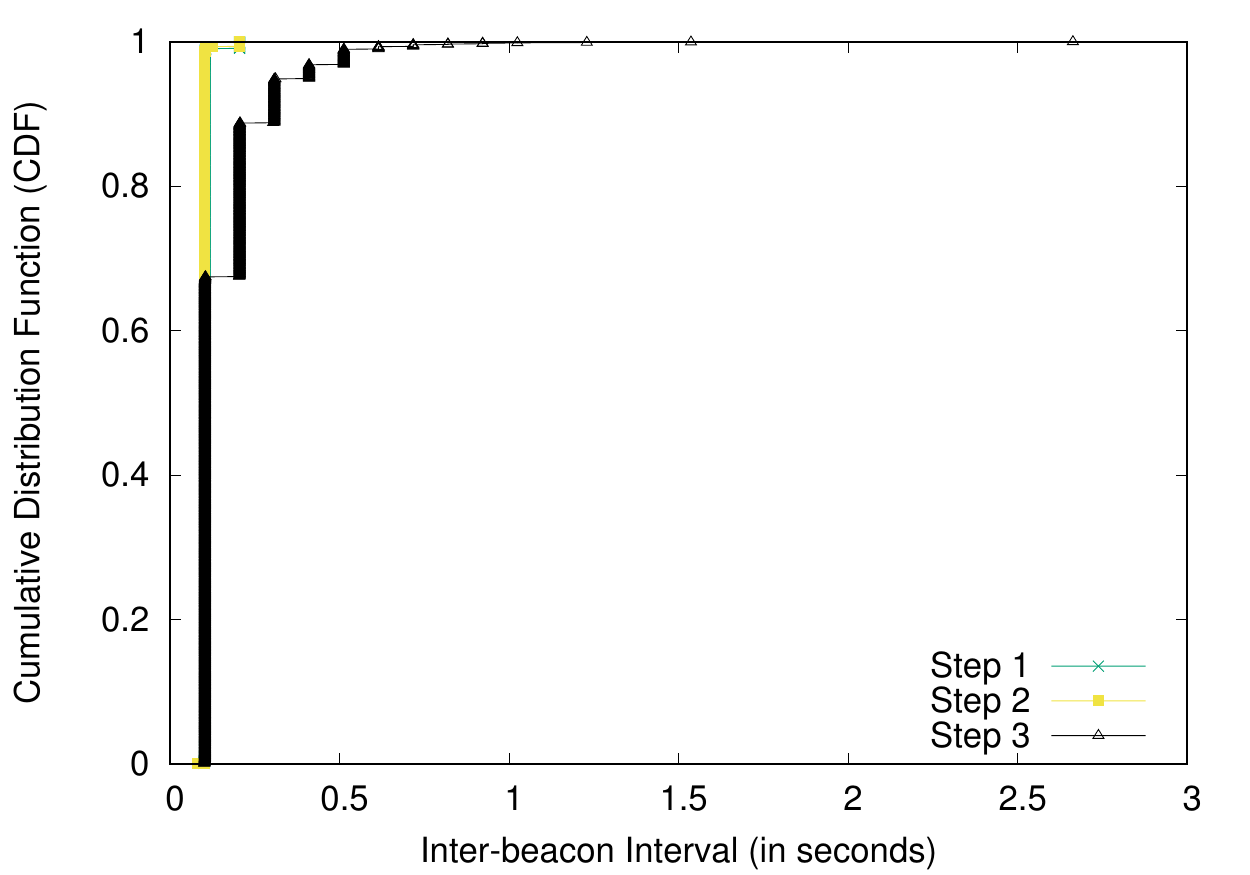}
  \vspace{-0.6cm}
  \caption{CDF of beacon interval for Steps 1, 2 and 3}
  \label{5b}
\end{figure}

Figure ~\ref{5b} shows the CDF of the inter-beacon interval when the two cells are separated by a distance of 17 feet. At this distance, there are no hidden-node issues to contend with. The inter-beacon time interval, should be close to 102.4 ms. But due to LTE-U interference, we see that in Step 3, this interval can be as large as 3 ms, which will have a significant effect on how long it takes for the LTE-U BS to detect the presence of the Wi-Fi AP. On the other hand, there is no perceptible difference in the inter-beacon times between a single Wi-Fi (Step 1) and two coexisting Wi-Fi APs (Step 2). Hence it is clear that LTE-U on an empty channel does not coexist with another Wi-Fi that wishes to share the channel in the same way that Wi-Fi coexists with Wi-Fi in the same situation with respect to beacon reception. We performed the experiment at different inter-cell distances and observed similar performance, which are not included due to space limitations.

\vspace{-0.1cm}
\subsection{Comparison with different LTE-U duty cycles and periods}
In the baseline comparison in the previous section, we only looked at the received beacons for a single duty cycle of 95\%. In this section we examine performance with different duty cycles and periods. We will consider the following cases:
\begin{itemize}
\item \textbf{Case A:} Wi-Fi/Wi-Fi Coexistence.
\item \textbf{Case B:} LTE-U/Wi-Fi Coexistence with 5 ms ON/5 ms OFF LTE-U duty cycle.
\item \textbf{Case C:} LTE-U/Wi-Fi Coexistence with 20 ms ON/20 ms OFF LTE-U duty cycle.
\item \textbf{Case D:} LTE-U/Wi-Fi Coexistence with 20 ms ON/1 ms OFF LTE-U duty cycle.
\item \textbf{Case E:} LTE-U/Wi-Fi Coexistence with 20 ms ON/5 ms OFF LTE-U duty cycle.
\end{itemize}
\begin{figure*}[!htb]
\begin{subfigure}[c]{0.5\textwidth}
  \centering
  \includegraphics[height=5cm,width=9cm]{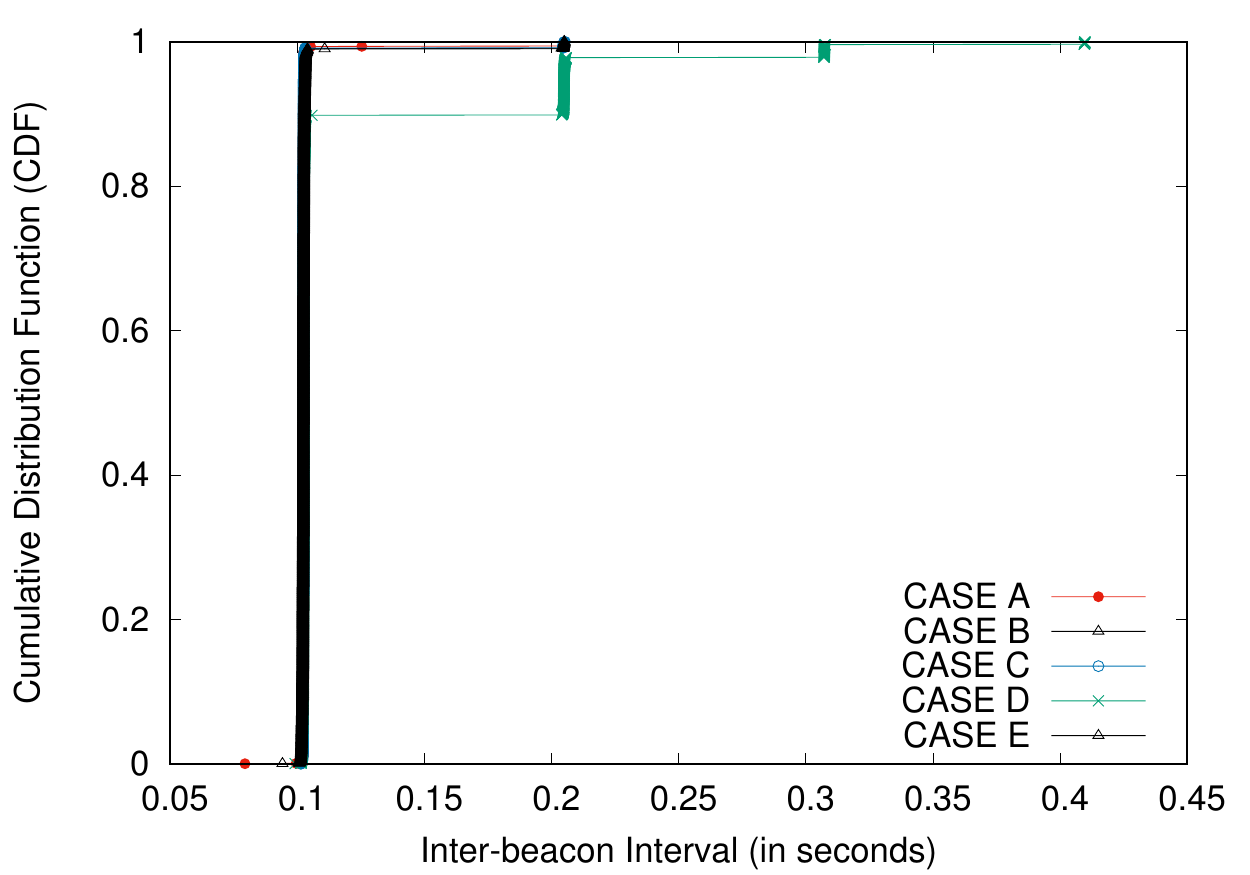}
\end{subfigure}
\begin{subfigure}[c]{0.5\textwidth}
  \centering
  \includegraphics[height=5cm,width=9cm]{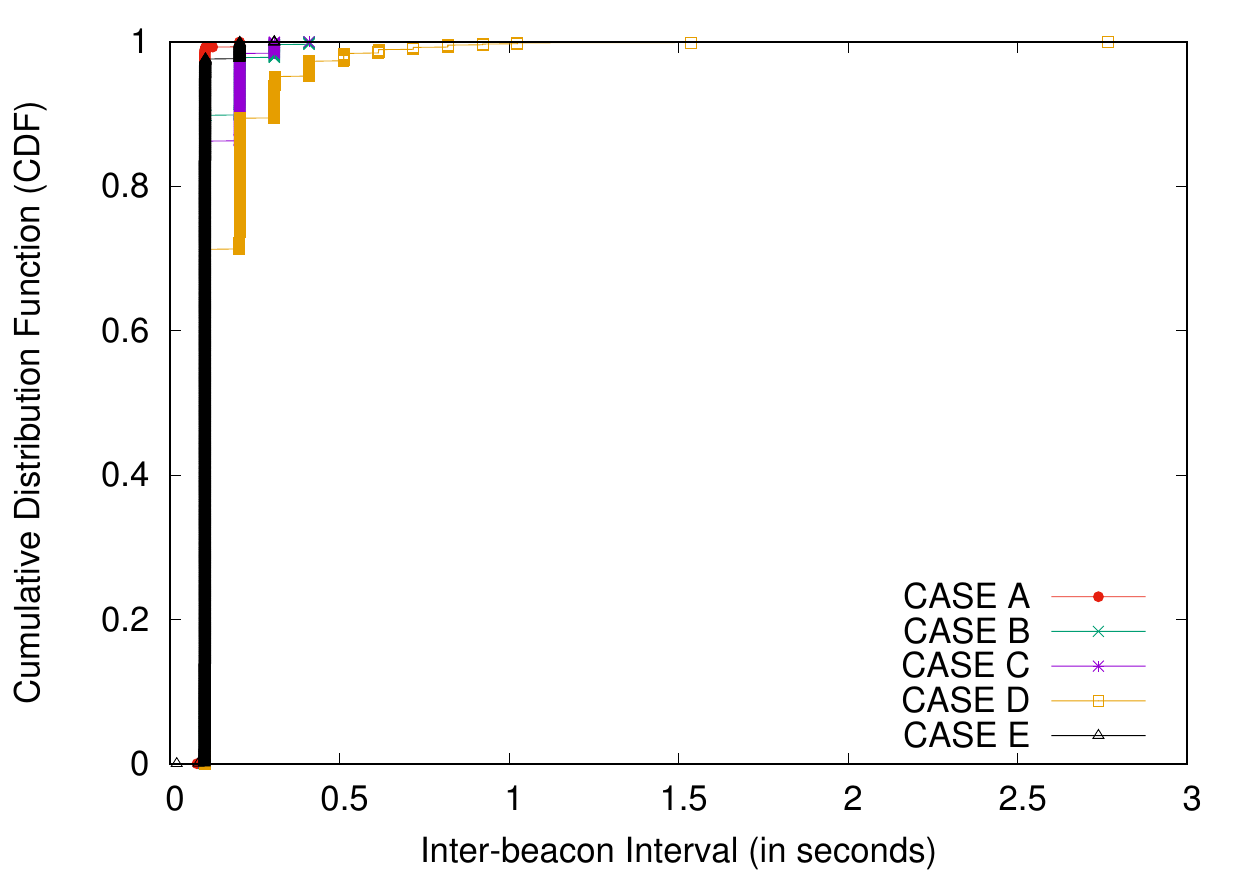} 
\end{subfigure}
 \vspace{-0.1cm}
\caption{(a) CDF of transmitted beacon intervals  and (b) CDF of received beacon intervals}
\vspace{-0.4cm}
 \label{be}
\end{figure*}
\vspace{-0.1cm}
\begin{figure}
\begin{center}
\includegraphics[height=5cm,width=9cm]{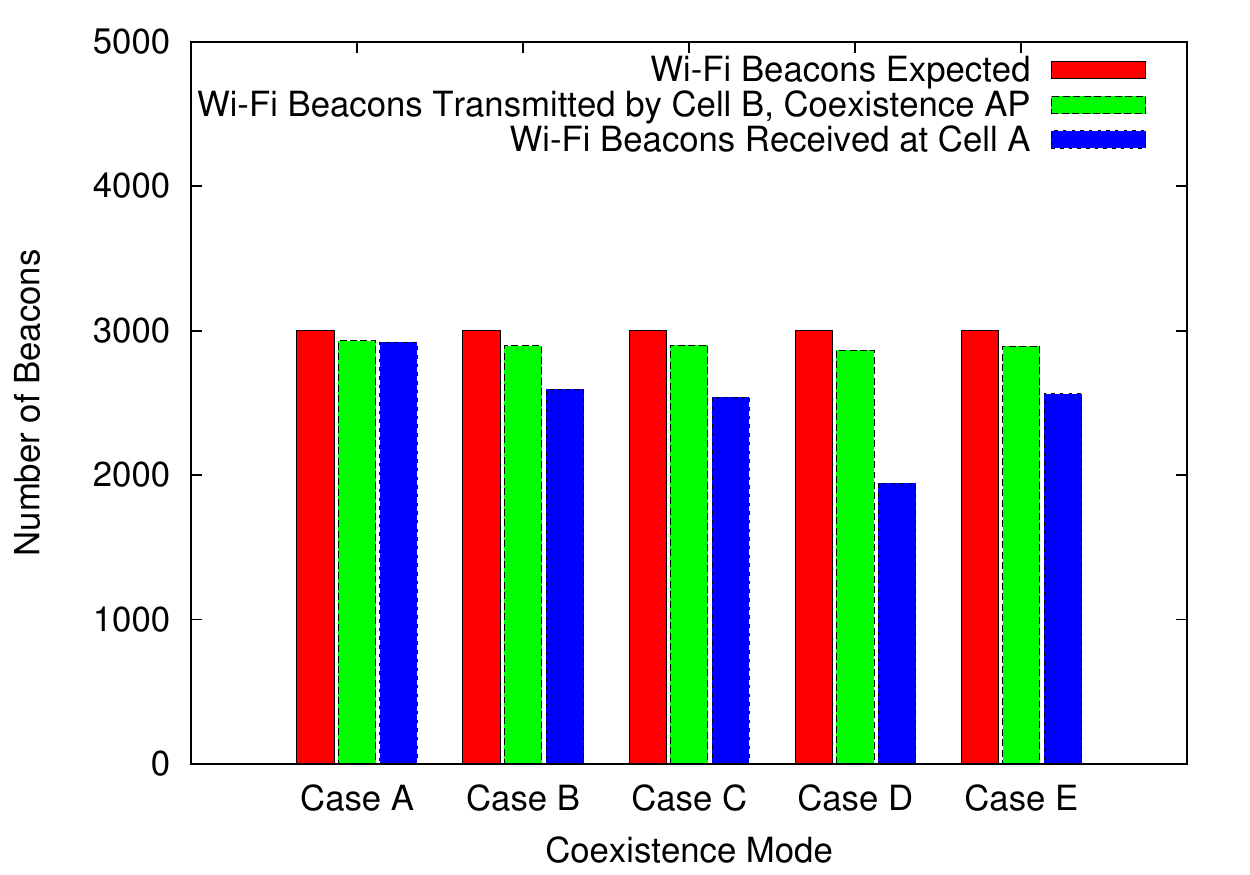}
\vspace{-0.5cm}
\caption{Beacons Reception at different ON/OFF modes.}
\label{breception}
\end{center}
\vspace{-0.7cm}
\end{figure}

In all cases above, beacon transmissions, receptions and inter-beacon interval measurements were made over a period of 5 mins, during which 3000 beacons should be transmitted. Beacons are transmitted by the $''$Coexistence AP$''$ in Cell B and are received at the LTE-U BS in Cell A, 17 feet away as shown in Figure ~\ref{tbed}. Figures.~\ref{be} (a) and~\ref{be} (b) show the CDF of transmitted beacon interval and received beacon interval respectively. We see that the transmitted and received beacon interval increases in Case D considerably, compared to the other scenarios, with the received beacon interval extending to seconds in some cases.

Figure ~\ref{breception} shows the total number of expected beacons, transmitted beacons and received beacons for the five cases above. In Case A, Wi-Fi/Wi-Fi coexistence, there is no appreciable drop in the number of beacons received. However in all of the other cases of LTE-U/Wi-Fi coexistence, there is a drop in the number of beacons received, even when the duty cycle is 50\% (Case B and Case C). In Case D, 20 ms ON/1 ms OFF, almost 1/3 of the transmitted beacon frames are not received at the LTE-U BS. Interestingly, Case E, with 80\% duty cycle has a similar beacon loss performance as Case B and C with 50\% duty cycle. Hence, it may be advisable for a LTE-U BS to use a duty cycle of no greater than 80\% in order to allow a Wi-Fi AP fair access to the medium. With 80\% duty cycle, the reduction in the number of beacons that are not received at the LTE-BS will also enable the LTE-U BS to react faster to the presence of Wi-Fi by reducing its duty cycle to the required 50\%.

\vspace{-0.3cm}
\section{Conclusions and Future Work}\label{p32}

In this paper, we performed theoretical analyses and extensive measurements with Wi-Fi and LTE-U in realistic deployments to understand the Wi-Fi beacon transmission/reception behavior when LTE-U operates on the same channel. We have shown very good agreement between the analyses and measurements. The motivation is to understand if LTE-U should be operating with its maximum allowed duty cycle of 95\% when it is operating on an empty channel. We find that if it does so, it will severely impact the ability of a Wi-Fi AP to share the channel since the beacon transmission/reception will be disrupted. Instead, if LTE-U scaled back the occupancy on an empty channel to 80\% (i.e. 20 ms ON/5 ms OFF), the Wi-Fi beacon loss reduces to an acceptable level and makes it easier for Wi-Fi to get on the air. 
Our work did not have a CSAT algorithm that would automatically detect Wi-Fi and scale back the duty cycle automatically. Future work will implement a realistic CSAT algorithm on the NI USRP and evaluate the performance.

This work was supported by NSF under grant CNS-1618920.

\nocite{*}
\bibliographystyle{ieeetr}
\bibliography{IEEEabrv,references}
\end{document}